\documentclass[10pt,letterpaper]{article}
\usepackage{opex3}
\usepackage{color}
\usepackage{bm}        
\usepackage{amssymb,amsmath,amsopn,amsfonts}
\usepackage{cite}

\newcommand{\ket}[1]{\ensuremath{\left| #1 \right\rangle}}
\newcommand{\bra}[1]{\ensuremath{\left\langle #1 \right|}}

\newcommand{\av}[1]{\ensuremath{\left\langle #1 \right\rangle}}

\def\eqref#1{Eq.~(\ref{#1})}
\def\eqsref#1{Eqs.~(\ref{#1})}
\newcommand{\vect}[1]{\bm{#1}}
\begin{document}

\title{Realization of mutually unbiased bases for a qubit with only one wave plate: Theory and experiment}

\author{Zhibo Hou,$^{1,2}$ Guoyong Xiang,$^{1,2,*}$ Daoyi Dong,$^{3}$ Chuan-Feng Li,$^{1,2}$ and Guang-Can Guo$^{1,2}$}
\address{$^{1}$\textit{Key Laboratory of Quantum Information,University of Science and Technology of China, CAS, Hefei 230026, China}\\
$^{2}$\textit{Synergetic Innovation Center of Quantum Information and Quantum Physics, University of Science and Technology of China, Hefei 230026, China}\\
$^{3}$\textit{School of Engineering and Information Technology, University of New South Wales, Canberra, ACT 2600, Australia}}
\email{$^*$gyxiang@ustc.edu.cn} 

\begin{abstract}
We consider the problem of implementing mutually unbiased bases (MUB) for a polarization qubit with only one wave plate, the minimum number of wave plates. We show that one wave plate is sufficient to realize two MUB as long as its phase shift (modulo $360^\circ$) ranges between $45^\circ$ and $315^\circ$. {It can realize} three MUB (a complete set of MUB for a qubit) if the phase shift of the wave plate is within $[111.5^\circ, 141.7^\circ]$ or its symmetric range with respect to 180$^\circ$. The systematic error of the realized MUB using a third-wave plate (TWP) with $120^\circ$ phase is calculated to be a half of that using the combination of a quarter-wave plate (QWP) and a half-wave plate (HWP). As experimental applications, TWPs are used in single-qubit and two-qubit quantum state tomography experiments and the results show a systematic error reduction by $50\%$. This technique not only saves one wave plate but also reduces the systematic error, which can be applied to quantum state tomography and other applications involving MUB. The proposed TWP may become a useful instrument in optical experiments, replacing multiple elements like QWP and HWP.
\end{abstract}

\ocis{(270.0270) Quantum optics; (230.0230) Optical devices.} 


\section{Introduction}\label{section:introduction}
Measurement bases with a special geometry structure  in Hilbert space, such as 2-designs \cite{Zaun11quantum,Rene04symmetric}, weighted 2-designs \cite{Roy07weighted}, tight frames \cite{Scot06tight} and Platonic solids \cite{Burg08choice},  have attracted wide attention in the community of quantum information science in recent years. A typical example of 2-designs is mutually unbiased bases {(MUB)} \cite{Durt10on,Woot89optimal}. The special geometry for MUB is the equal Hilbert-Schmidt overlap between any two projective operators corresponding to two MUB. Hence, measurement results of {one operator} provide no information about the measurement results of its MUB at all. Due to this geometry, MUB have found applications from quantum key distribution \cite{Benn84quantum,Gisi02quantum,Niel00quantum}, entropy uncertainty relations \cite{Wehn10entropic,Wu09entropic,Ball07entropic} to quantum state estimation \cite{Woot89optimal,Lima11experimental,Zhu12quantum,Adam10improving,Qi13quantum}.

A complete set of MUB in a multi-qubit system involves nonlocal measurements that are difficult to be realized in the laboratory. In many scenarios such as quantum state tomography \cite{Pari04quantum,Fan07quantum,Cram10efficient,Alte05photonic}, the tensor product of a complete set of {single-qubit MUB is a preferred choice as the measurement bases for multi-qubit systems} \cite{Burg08choice,Qi13quantum}. Therefore, it is important to implement MUB for a single qubit in experiments \cite{Mahl13adaptive,Krav13experimental,Salv13full}.

In the context of polarization optics, a specific type of polarization transformation poses constraints on the minimum number of optical elements that must be used. Three wave plates are the minimum number to realize any SU(2) polarization transformations \cite{Simo90minimal} and the visual tool kit can be found in \cite{Simo12hamilton}. {As for} transformations between a given pair of nonorthogonal polarization states, two QWPs are enough (see a sketched proof in \cite{Dama04polarization} and a recent constructive demonstration by Zela in \cite{Zela12two}). When the initial state is linearly polarized, {a quarter-wave plate (QWP) and a half-wave plate (HWP)} can realize any polarization state \cite{Dama04polarization}. The combination of QWP and HWP is used in most of current optical experiments \cite{Jame01measurement} where the linear polarization is generated by a polarizing beam splitter (PBS). In the setting composed of a QWP, a HWP and a PBS, the PBS acts as $\sigma_z$ and by adjusting the optic axis angles of the QWP and HWP (a two-parameter setting), the Bloch vector of $\sigma_z$ can be unitarily rotated to any point on the unit Bloch sphere \cite{Dama04polarization}. However, when the problem is restricted to realize MUB rather than arbitrary bases, is one wave plate sufficient to transform the initial $\sigma_z$ to a set of MUB, especially a complete set of three MUB? If the answer is positive, not only one wave plate is saved but also the systematic error \cite{Jame01measurement,Ross12imperfect} can be expected to decrease as the parameter uncertainties of the devices (i.e., wave plates in our case) are the sources of systematic error. Intuitively, the less measurement devices, the smaller the systematic error. So far, people know how to construct two MUB with one HWP or one QWP. 
For example, one HWP itself can be used to realize $\sigma_z$ and $\sigma_x$ by setting its rotation angles as $0^\circ$ and $22.5^\circ$; one QWP can be used to realize $\sigma_z$ and $\sigma_y$ by setting its rotation angles as $0^\circ$ and $45^\circ$. One wave plate is also used to perform Fourier transform tomography \cite{Moha13fourier} and the optimal phase of the wave plate is numerically calculated in \cite{Moha14optimization}. As to a complete set of three MUB, this problem is barely considered. In this paper, we show that it is indeed feasible to employ only one wave plate to realize a complete set of single-qubit MUB with reduced systematic error.

Here is the organization for the rest of the paper. In section \ref{section:theory}, the conditions to realize two MUB and three MUB using only one wave plate are considered, respectively. Section \ref{section:systematic error of MUB} calculates the systematic error in the realization of MUB with only one wave plate. In section \ref{section:experiments}, third-wave plates are used to perform MUB measurements in both single qubit and two-qubit tomography experiments and the results demonstrate an error reduction by $50\%$. Section \ref{section:conclusion} concludes the paper.

\section{Theory}\label{section:theory}

\subsection{{ Transforming} polarization states with QWP-HWP setting}
In optical experiments, arbitrary projective measurements on polarization qubits are often implemented by a QWP-HWP setting; see Fig.~\ref{configuration}(a). Define $|H\rangle=(1, 0)^T, |V\rangle=(0, 1)^T$ and their corresponding eigenvalues are $\pm1$, where $^T$ denotes transpose. Then the PBS acts as $\sigma_z$ or $(0, 0, 1)^T$ in the Bloch representation. This setup transforms \ket{H} to
\begin{equation}\label{psi}
  \ket{\psi}=U(q,\frac{\pi}{2})U(h,\pi)\ket{H}=\left(
                                                                \begin{array}{c}
\cos q\cos (q-2h)+i\sin q\sin (q-2h) \\
\sin q\cos (q-2h)-i\cos q\sin (q-2h) \\
                                                                \end{array}
                                                              \right),
\end{equation}
where $q,h$ are{, respectively, }the optic axis angles of QWP and HWP deviated from the horizontal direction, and $U$ is a unitary transformation operator on polarization by a phase plate with phase $\delta$ and rotation angle $\theta$,
\begin{equation}\label{unitary}
  U(\theta,\delta)=\left(
      \begin{array}{cc}
        \cos^2\theta+e^{i\delta}\sin^2\theta & \frac{1}{2}(1-e^{i\delta})\sin 2\theta \\
        \frac{1}{2}(1-e^{i\delta})\sin 2\theta & \sin^2\theta+e^{i\delta}\cos^2\theta \\
      \end{array}
    \right).
\end{equation}
{Or equivalently this setup transforms the original operator $\sigma_z$ to the Pauli operator $\vect{r}(q,h)\cdot\vect{\sigma}$ with \begin{equation}\label{r:QWP-HWP}
\vect{r}(q,h)=\left(\sin 2q\cos (4h-2q),\sin (4h-2q),\cos 2q\cos (4h-2q)\right)^T
\end{equation}
and $ \vect{\sigma}=(\sigma_x, \sigma_y, \sigma_z)$, where $\sigma_x$, $\sigma_y$ and $\sigma_z$ are three Pauli operators. 

 Two variables of rotation angles $q$ and $h$ correspond to a plane and can rotate the initial Bloch vector $(0, 0, 1)^T$ to any direction or position on the unit Bloch sphere. {Thus,} this QWP-HWP setting is used to realize any one qubit projective measurement basis. For example, the most popular set of MUB is composed of Bloch vectors $(1, 0, 0)^T$, $(0, 1, 0)^T$ and $(0, 0, 1)^T$ (corresponding to three Pauli operators) which are realized by choosing $q=45^\circ,0^\circ,0^\circ$ and $h=22.5^\circ,22.5^\circ,0^\circ$  correspondingly.

In order to realize MUB for a qubit, only two or three specific Bloch vectors need to be implemented and it is not necessary to cover the Bloch sphere with two rotation angles of two wave plates. We will show below that one wave plate with appropriate phases is sufficient to realize single-qubit MUB.

\subsection{Realizing two mutually unbiased bases with one wave plate}
Two { sets of} orthogonal bases $\{|\psi_i^1\rangle\}^{d}_{i=1}$ and $\{|\psi_j^2\rangle\}^{d}_{j=1}$ in a $d$-dimensional Hilbert space are mutually unbiased if ${|\langle\psi_i^1|\psi_j^2\rangle|}^2=\frac{1}{d}$ for any $i$ and $j$. There are at most $d+1$ bases with every two of them mutually unbiased, which are called a complete set of MUB. So far, how to construct a complete set of MUB is known only in systems with dimensions which are powers of primes \cite{Durt10on,Woot89optimal}. {For general cases}, the existence of such a set is still an open problem even in the simple case of $d=6$ \cite{Durt10on,Brie09constructing}. For a qubit with dimension $d=2$, the bases $\{|\psi_i^j\rangle\}^2_{i=1}$ are the eigenvectors of a unit Pauli operator $\vect{r}^j\cdot\vect{\sigma}$ with $\|\vect{r}^j\|=1$. In the Bloch representation, the $j$th basis is directly related to its Bloch vector $\vect{r}^j$ as $|\psi_i^j\rangle\langle\psi_i^j|=\frac{I\pm\vect{r}^j\cdot\vect{\sigma}}{2}$ and $\pm$ corresponds to $i=1, 2$. Thus, the requirement for two bases to be mutually unbiased is that their Bloch vectors are normal to each other
\begin{equation}\label{mub condition}
  \vect{r}^ 1\cdot\vect{r}^ 2=0.
\end{equation}

\begin{figure}
\center{\includegraphics[scale=0.47]{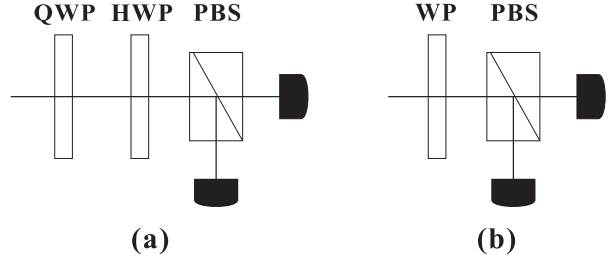}}
\caption{\label{configuration}{MUB measurement with two different settings: (a) {MUB} measurement with QWP-HWP setting consisting of a QWP, a HWP and a PBS. (b) MUB measurement with a wave plate (WP) and a PBS.}}
\end{figure}

In Fig.~\ref{configuration}(b), a wave plate (WP) with a phase difference $\delta$ between ordinary (o) and extraordinary (e) components is placed in front of a PBS. Its unitary transformation operator on polarization is expressed in \eqref{unitary}
where $\theta$ is the deviation angle of the optic axis from the horizontal direction. This wave plate with rotation angle $\theta$ transforms the initial Bloch vector $\vect{r}_0=(0, 0, 1)^T$ into
\begin{equation}\label{bloch vector r}
  \vect{r}(\theta)=\begin{array}{ccc}
    (\sin^2\frac{\delta}{2}\sin 4\theta, & -\sin \delta\sin 2\theta, & 1-2\sin^2\frac{\delta}{2}\sin^22\theta)^T.
    \end{array}
\end{equation}
From \eqref{mub condition}, this setting with one wave plate can realize two MUB if there exist $\theta_1$ and $\theta_2$ such that their Bloch vectors are normal to each other, i.e., $\vect{r}(\theta_1)\cdot\vect{r}(\theta_2)=0$. The existence of $\theta_1$ and $\theta_2$ obviously depends on $\delta$. One trivial example is $\delta=0^\circ$ where the wave plate does not affect the initial polarization and $\vect{r}(\theta_1)\cdot\vect{r}(\theta_2)=1$ for any $\theta_1, \theta_2$. Another example is $\delta=180^\circ$. The choice of $\theta_1=0^\circ$ and $\theta_2=22.5^\circ$ makes $\vect{r}(\theta_1)=(0,0,1)^T$, $\vect{r}(\theta_2)=(1,0,0)^T$, and thus $\vect{r}(\theta_1)\cdot\vect{r}(\theta_2)=0$. The maximum of $\vect{r}(\theta_1)\cdot\vect{r}(\theta_2)$ is 1, which can be achieved at $\theta_1=\theta_2$. {Because $\vect{r}(\theta_1)\cdot\vect{r}(\theta_2)$ is a continuous function and its maximum is positive, there exist $\theta_1$ and $\theta_2$ to make $\vect{r}(\theta_1)\cdot\vect{r}(\theta_2)=0$ if the minimum of $\vect{r}(\theta_1)\cdot\vect{r}(\theta_2)$ is negative or zero.} This problem is then converted into finding the minimum of $\vect{r}(\theta_1)\cdot\vect{r}(\theta_2)$. From \eqref{bloch vector r}, the first two elements of $\vect{r}$ are odd functions of $\theta$ and the third term is an even function of $\theta$. That is, $r_1(-\theta)=-r_1(\theta), r_2(-\theta)=-r_2(\theta)$ and $r_3(-\theta)=r_3(\theta)$. From \eqref{bloch vector r}, the minimum and maximum of $r_3(\theta)$ is $\cos {\delta}$ and 1. If $\cos {\delta}$ is non-positive, then there exists $\theta_1=\frac{1}{2}{\arcsin}(\frac{\sqrt{2}}{2}{\csc}\frac{\delta}{2})$ that makes $r_3(\theta_1)=0$ and we choose $\theta_2=-\theta_1$ to make $\vect{r}(\theta_1)=-\vect{r}(\theta_2)$. We obtain the minimum $\vect{r}(\theta_1)\cdot\vect{r}(\theta_2)=-1$. Thus, in this case, two MUB can be realized such as $\theta_1=\frac{1}{2}{\arcsin}(\frac{\sqrt{2}}{2}{\csc}\frac{\delta}{2})$ and $\theta_2=0^\circ$. If $\cos {\delta}$ is positive, the minimum of $\vect{r}(\theta_1)\cdot\vect{r}(\theta_2)$ is $\cos 2\delta$ obtained at $\theta_1=-\theta_2=45^\circ$. Thus, the requirements that the minimum of $\vect{r}(\theta_1)\cdot\vect{r}(\theta_2)$ is non-positive in these two cases together give $45^\circ\leq\delta\leq315^\circ$. As long as the phase $\delta$ (modula $360^\circ$) of the wave plate is within $[45^\circ,315^\circ]$, it can be used to realize at least two MUB. As there are two parameters and one equation, the solutions of \eqref{mub condition} form a line. One solution is $\theta_1=\frac{1}{2}{\arcsin}(\frac{\sqrt{2-\sqrt{2}}}{2}{\csc}\frac{\delta}{2})$ and $\theta_2=-\theta_1$, which make $r_3=\frac{\sqrt{2}}{2}$. The Bloch vectors of the two MUB are axially symmetric about the initial state $(0,0,1)^T$, and the angles between MUB and $(0,0,1)^T$ are $45^\circ$.

\subsection{Realizing a complete set of MUB with one wave plate}

\begin{figure}
\center{\includegraphics[scale=0.65]{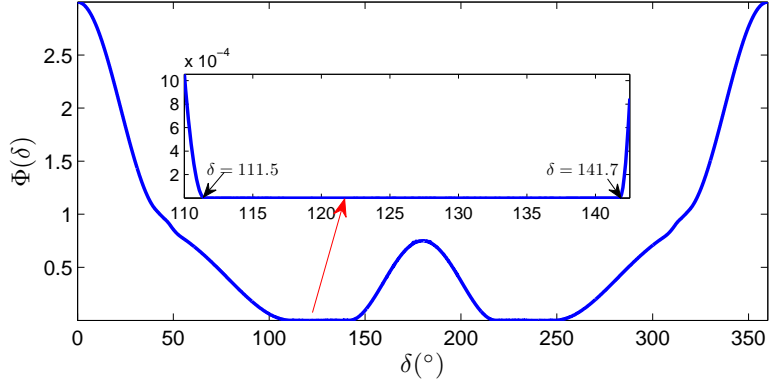}}
\caption{\label{delta_range}{Frame potential $\Phi(\delta)$ with respect to $\delta$. The frame potential is symmetric about $180^\circ$. The inset is the amplified interval of $\delta$ where the frame potential vanishes to zero. Only if $\delta$ is in the range $[111.5^\circ,141.7^\circ]$ or its symmetric range about $180^\circ$ {can a complete set of MUB be realized by one wave plate}.}}
\end{figure}

Three rotation angles $\theta_i, i=1, 2, 3,$ are chosen to realize three MUB, which consist of a complete set of MUB for a qubit,
\begin{equation}\label{3 rotation angles}
  \vect{r}(\theta_i)\cdot\vect{r}(\theta_j)=0,
\end{equation}
where $i, j=1, 2, 3$ and $i\neq j$. For a complete set of MUB, it is much more complicated and difficult to obtain theoretical solutions. Hence, the problem is investigated numerically in this paper. Borrowed from frame theory \cite{Rene04symmetric,Bene03finite}, the frame potential of a wave plate is defined as
\begin{equation}\label{frame potential}
\Phi(\delta)=\mathop\text{min}\limits_{\theta_1, \theta_2, \theta_3} [\vect{r}(\theta_1)\cdot\vect{r}(\theta_2)]^2+[\vect{r}(\theta_2)\cdot\vect{r}(\theta_3)]^2+[\vect{r}(\theta_1)\cdot\vect{r}(\theta_3)]^2.
\end{equation}
The solution of Eq.~(\ref{3 rotation angles}) exists only if the phase $\delta$ of the wave plate makes the frame potential \eqref{frame potential} vanishing. The frame potential is numerically computed by a MATLAB solver {\it{lsqnonlin}}, which is intended to solve nonlinear least-squares problems. One hundred different initial points of $\theta_1,\theta_2$ and $\theta_3$ are taken for us to avoid local minimum and find all the possible solutions. The numerical result of the frame potential for different phases is shown in Fig.~\ref{delta_range}.
 From Eq.~(\ref{bloch vector r}), $\Phi(\delta)$ has a period of 360$^\circ$. Thus, we only consider $0^\circ\leq\delta<360^\circ$. Since $\vect{r}(\theta_i, \delta)\cdot\vect{r}(\theta_j, \delta)=\vect{r}(\theta_i, 360^\circ-\delta)\cdot\vect{r}(\theta_j, 360^\circ-\delta)$, we {have} $\Phi(\delta)=\Phi(360^\circ-\delta)$, meaning that the frame potential $\Phi(\delta)$ is symmetric about $\delta=180^\circ$.

\begin{figure}
\center{\includegraphics[scale=0.6]{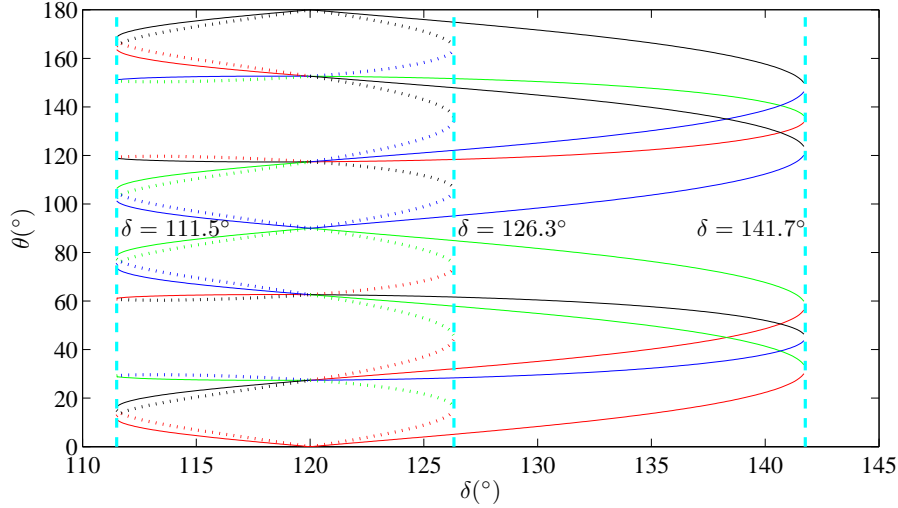}}
\caption{\label{all_angles}{The numerical solutions of rotation angles in Eq.~(\ref{3 rotation angles}) with respect to the phase $\delta$. There are two classes of solutions with $\delta$ between $111.5^\circ$ and $126.3^\circ$, represented by solid and dotted lines, respectively, {and only one class of solutions in $(126.3^\circ,141.7^\circ]$}. Each class of solutions contains four complete sets of MUB, denoted by four different colors.}}
\end{figure}


With $\delta$ between $111.5^\circ$ and $141.7^\circ$, the frame potential is zero and the solutions of Eq.~(\ref{3 rotation angles}) are shown in Fig.~\ref{all_angles}. From \eqref{bloch vector r}, $\vect{r}(\theta)=\vect{r}(\theta+180^\circ)$, we only consider rotation angles within $[0^\circ,180^\circ)$. {} \eqref{bloch vector r} also shows $\vect{r}(\theta_i)\cdot\vect{r}(\theta_j)=\vect{r}(90^\circ\pm\theta_i)\cdot\vect{r}(90^\circ\pm\theta_j)=\vect{r}(180^\circ-\theta_i)\cdot\vect{r}(180^\circ-\theta_j)$. That is to say, if $\theta_1,\theta_2,\theta_3$ {(red in Fig.~\ref{all_angles})} is the solution, the modules of $90^\circ\pm\theta_i$ and $180^\circ-\theta_i$ $(i=1,2,3)$ by $180^\circ$ {(represented as blue, green and black in Fig.~\ref{all_angles} correspondingly)} are also the solutions. Thus, all the four sets of MUB are considered as one class of solutions. 

As shown in Fig.~\ref{all_angles}, solutions from different sets of MUB in the same class intersect around $\delta=120^\circ, 126.3^\circ$ and $141.7^\circ$. Using the symmetries represented by the colors, the four variables (i.e. $\delta,\theta_1,\theta_2$ and $\theta_3$) reduce to two and we can theoretically calculate the solutions at these intersections. The phases at these intersections are also rigourously found to be  $\delta=120^\circ, 126.32^\circ$ and $141.76^\circ$ (see Appendix \ref{section:APPEND-120} and \ref{section:APPEND-126.3 and 141.7}). Here we compare our results with those in \cite{Moha14optimization}. The optimal phase was numerically calculated in \cite{Moha14optimization} to be $7\pi/10$ (i.e. $126^\circ$). The figure of merit times the total number of counts in \cite{Moha14optimization} at this optimal phase equals 10.03, which is very close to 10, the bound achieved by MUB. This phase falls within our calculated range $[111.5^\circ,141.7^\circ]$. The reason why they only found one phase rather than an available interval of phases and the optimal performance at this phase was slightly worse than the bound is their restriction of six equally spaced rotation angles.

In the special case of $\delta_t=120^\circ$, called third-wave plate (TWP), the Bloch vector is
\begin{equation}\label{r:TWP}
\vect{r}({t})=\left(\frac{3}{4}\sin 4{t},-\frac{\sqrt{3}}{2}\sin 2{t},\frac{3}{4}\cos 4{t}+\frac{1}{4}\right)^T,
\end{equation}
where $t$ is the rotation angle of the optic axis of TWP deviated from horizontal direction. From Appendix \ref{section:APPEND-120}, solutions of \eqref{3 rotation angles} are $\vect{r}(0^\circ)=\vect{r}(90^\circ)=\vect{r}(180^\circ)=(0, 0, 1)^T, \vect{r}({t}_0)=-\vect{r}(180^\circ-{t}_0)=\frac{1}{\sqrt{2}}(1, -1, 0)^T, \vect{r}({t}_0+90^\circ)=-\vect{r}(90^\circ-{t}_0)=\frac{1}{\sqrt{2}}(1, 1, 0)^T$, where ${t}_0=\frac{1}{4}\arccos(-\frac{1}{3})\approx 27.37^\circ$.


\section{Systematic error in the realization of MUB in one wave plate setting}\label{section:systematic error of MUB}

Imperfect measurement devices are the main sources of the systematic error in the realization of MUB. Here we consider the systematic error due to the parameter uncertainties of wave plates. The realized bases are denoted by their Bloch vectors as $\vect{r}(\delta, \theta)$, where $\delta$ is the real phase of the wave plate in the one wave plate setting in Fig.~\ref{configuration}(b). The systematic error in the realization of $\vect{r}$ is
\begin{equation}\label{one operator error}
  (\Delta{\vect{r}})^2=\sum_{\xi}{\|\vect{r}_{\xi}\|^2(\Delta{{\xi}})^2},
\end{equation}
where $\vect{r}_{\xi}=\frac{\partial{\vect{r}}}{\partial{{\xi}}}$, ${\xi}=\delta, \theta$. From \eqref{bloch vector r},
 \begin{equation}\label{EQ: r derivative TWP}
 \begin{aligned}
   \|\vect{r}_{\delta}\|^2&=\sin^22{\theta}=\frac{(r_2)^2}{\sin^2\delta},\\
   \|\vect{r}_{\theta}\|^2&=16\sin^2\frac{\delta}{2}-4\sin^2\delta\sin ^{2}2{\theta}=16\sin^2\frac{\delta}{2}-4(r_2)^2.
 \end{aligned}
 \end{equation}
For a complete set of MUB ($\vect{r}^j, j=1,2,3$), the systematic error sums up to
\begin{equation}\label{EQ:three MUB error}
\begin{aligned}
\epsilon^2&=\sum_{j=1}^3(\Delta{\vect{r}^j})^2=\sum_{j=1}^3{\sum_{\xi}{\|\vect{r}^j_{\xi}\|^2(\Delta{{\xi}})^2}}=\sum_{{\xi}}({\epsilon_{\xi}})^2(\Delta {\xi})^2,\\
({\epsilon_{\xi}})^2&=\sum_{j=1}^{3}\|\vect{r}^j_{\xi}\|^2.
\end{aligned}
\end{equation}
As $\vect{r}^j$ is orthogonal to each other, from \eqref{EQ: r derivative TWP},
\begin{eqnarray*}
  ({\epsilon_{\delta}})^2 &=& \sum_{j=1}^{3}\frac{(r_2^j)^2}{\sin^2\delta}=\frac{1}{\sin^2\delta}, \\
  ({\epsilon_\theta})^2 &=& \sum_{j=1}^{3}16\sin^2\frac{\delta}{2}-4(r_2^j)^2=48\sin^2\frac{\delta}{2}-4.
\end{eqnarray*}
Thus, the systematic error in the one wave plate setting is
 \begin{equation}\label{EQ:systematic error with one wave plate}
 \epsilon^2=\frac{1}{\sin^2\delta}(\Delta\delta)^2+(48\sin^2\frac{\delta}{2}-4)(\Delta\theta)^2.
\end{equation}
As $\epsilon^2$ in \eqref{EQ:systematic error with one wave plate} is an increasing function of $\delta$ in the interval $[111.5^\circ, 141.7^\circ]$, the minimum and maximum systematic error in the realization of three MUB with one wave plate setting is $1.16(\Delta\delta)^2+28.80(\Delta\theta)^2$ and $2.60(\Delta\delta)^2+38.83(\Delta\theta)^2$, achieved at $\delta=111.5^\circ$ and $\delta=141.7^\circ$. For a third-wave plate with $\delta_t=120^\circ$,
\begin{equation}\label{EQ: systematic error-TWP}
\epsilon^2=1.33(\Delta{\delta_t})^2+32(\Delta{t})^2.
\end{equation}
Under the assumption that $(\Delta \delta_h)^2=(\Delta \delta_q)^2=(\Delta \delta_t)^2=(\Delta \delta)^2$ and $(\Delta h)^2=(\Delta q)^2=(\Delta t)^2=(\Delta \theta)^2$, the systematic error in the realization of MUB is $1.33(\Delta{\delta})^2+32(\Delta\theta)^2$ in the TWP setting and averaged as $(2.5\Delta\delta)^2+68(\Delta\theta)^2$ in the QWP-HWP setting in \eqref{EQ: systematic errors-QH}. Thus, the TWP setting outperforms the QWP-HWP setting by about a factor of two.

Measurements based on single-qubit MUB are preferable choices in quantum state tomography. In qubit state estimation, a complete set of single-qubit MUB is used to extract information of the qubit optimally. In multi-qubit quantum state tomography, the product measurements of single-qubit MUB on each photon are used to reduce estimation error due to statistical fluctuation. When the copies of states $\rho$ are infinite, the estimated state $\hat{\rho}$ based on the measurement data should be the same as the real state $\rho$. However, since single-qubit MUB are imperfectly realized, the estimated state $\hat{\rho}$ no longer converges to the real state $\rho$ and $\text{tr}(\hat{\rho}-\rho)^2$ is defined as the systematic error in state estimation. Generally, $\text{tr}(\hat{\rho}-\rho)^2$ depends on $\rho$. Averaged over unitarily equivalent states, $\av{\text{tr}(\hat{\rho}-\rho)^2}$ in both single-qubit and multi-qubit state estimation is proportionate to the systematic error of the realized bases \cite{hou_prepare}. As the systematic error in the realization of multi-qubit product bases is the sum over the systematic error of single MUB for each qubit, the systematic error in the realized multi-qubit product bases with one wave plate for each qubit is still a half of that with the QWP-HWP combination. 
This systematic error reduction effect in quantum state tomography is experimentally verified in both single-qubit and two-qubit tomography experiments in the next section.

\section{Quantum state tomography experiments with third-wave plates}\label{section:experiments}
\subsection{Qubit tomography experiments}

\begin{figure}
\center{\includegraphics[scale=0.5]{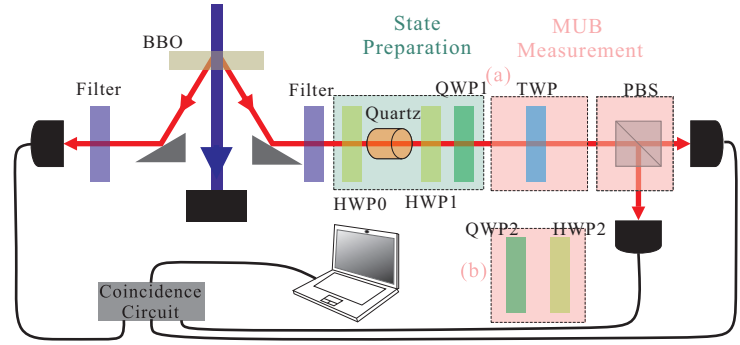}}
\caption{\label{third_wave_plate_experimental_setup}{Experimental setup for qubit tomography. The apparatus consists of two parts: state preparation (green) and MUB measurement (pink). The MUB measurement part consists of a polarizing beam splitter and a wave-plate combination which has two choices: (a) TWP and (b) QWP-HWP combination.}}
\end{figure}

\begin{figure}
\center{\includegraphics[scale=0.55]{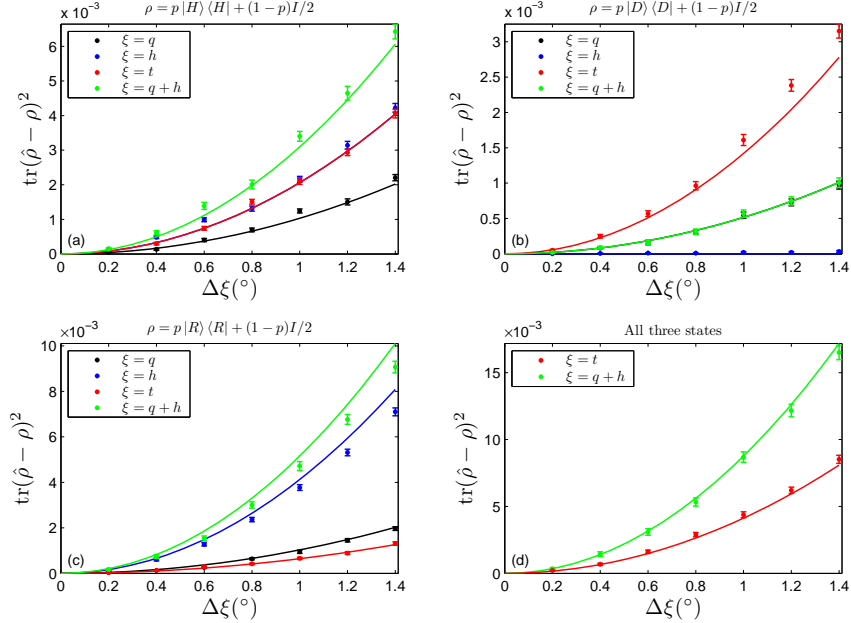}}
\caption{\label{error_1_qubit}{Systematic error in TWP and QWP-HWP setting for qubit tomography. The dependence of the systematic error in the TWP setting (Fig.~\ref{third_wave_plate_experimental_setup}(a)) and the QWP-HWP setting (Fig.~\ref{third_wave_plate_experimental_setup}(b)) is experimentally measured with respect to the angle errors of TWP (red), QWP (black) and HWP (blue) for three states in {Figs.~\ref{error_1_qubit}(a)--\ref{error_1_qubit}(c)} at $p=0.92$. The total systematic error (green) in the QWP-HWP setting is the sum of that due to QWP and HWP. The experimental results denoted as dots coincide with the theoretical calculations (solid lines). Fig.~\ref{error_1_qubit}(d) plots the total systematic error for all these three states in the two settings and shows that the TWP setting beats the QWP-HWP setting by a factor of two. Error bars are the standard deviation of 100 trials in Monte Carlo simulation with binomial distribution of counting statistics. }}
\end{figure}

The experimental setup, shown in Fig.~\ref{third_wave_plate_experimental_setup}, includes two parts: state preparation and MUB measurement. A 40 mW, V-polarized beam at 404 nm from a semiconductor laser pumps a type I phase-matched $\beta$-barium borate (BBO) crystal. After the spontaneous parametric down-conversion (SPDC) process, a pair of 808 nm H-polarized photons are created. One photon passes through a 3 nm interference filter and is detected by a single photon detector to herald the presence of its twin photon. The quantum state of the heralded photon is prepared by HWP0, HWP1, QWP1 and a 770$\lambda$ quartz crystal which is much larger than the coherence length of about 270$\lambda$ with $\lambda=808$ nm. HWP0 with rotation angle $h_0$ and the quartz crystal with optic axis aligned horizontally together prepare the quantum state $\rho$ wih Bloch vector $\vect{s}=\cos4h_0(0, 0, 1)^T$; HWP1 and QWP1 can transform $\vect{s}$ to arbitrary direction. This part is capable of preparing arbitrary qubit state. In the MUB measurement part, a complete set of MUB is performed with two methods: a TWP with rotation angles set as $0^\circ$, $27.37^\circ$ and $117.37^\circ$ in Fig.~\ref{third_wave_plate_experimental_setup}(a); the conventional QWP-HWP setting with rotation angles set as $(45^\circ, 22.5^\circ)$, $(0^\circ, 22.5^\circ)$ and $(0^\circ, 0^\circ)$ in Fig.~\ref{third_wave_plate_experimental_setup}(b).

As the systematic error in quantum state tomography experiments depends on the state to be measured, we prepared three states $\rho=p\ket{\phi}\bra{\phi}+(1-p){I}/{2}$ at $p=0.92$ (by setting HWP0's rotation angle $h_0=5.8^\circ$) and $\ket{\phi}=\ket{H}, \ket{D}$ and $\ket{R}$ {with a fidelity of more than 0.998.} The theoretical systematic error ${\text{tr}(\hat{\rho}-\rho)^2}$ in the estimation of these three states  is derived in Appendix~\ref{section:APPEND-qubit estimation systeamtic error} as $[8(\Delta~h)^2+4(\Delta~q)+0.25(\Delta\delta_h)^2]p^2$, $[2(\Delta~q)+0.25(\Delta\delta_h)^2+0.5(\Delta\delta_q)^2]p^2$ and $[16(\Delta~h)^2+4(\Delta~q)]p^2$ in the QWP-HWP setting and $[8(\Delta{t})^2+\frac{1}{3}(\Delta{\delta_t})^2]p^2$, $[\frac{11}{2}(\Delta{t})+\frac{1}{6}(\Delta{\delta_t})^2]p^2$ and $[\frac{5}{2}(\Delta{t})^2+\frac{1}{6}(\Delta{\delta_t})]p^2$ in the TWP setting. Under the assumption that $(\Delta \delta_h)^2=(\Delta \delta_q)^2=(\Delta \delta_t)^2=(\Delta \delta)^2$ and $(\Delta h)^2=(\Delta q)^2=(\Delta {t})^2=(\Delta\theta)^2$, the total systematic error of these three states is $[(\Delta\delta)^2+34(\Delta\theta)^2]p^2$ in the QWP-HWP setting and $[\frac{2}{3}(\Delta\delta)^2+16(\Delta\theta)^2]p^2$ in one wave plate setting. That is, the systematic error in realizing the corresponding MUB using the TWP setting is around a half of that using the QWP-HWP setting.

As phase errors are determined by the manufacture and wavelength, without a variable wavelength we can only experimentally verify the relationship between the systematic error of the estimated state and the angle errors of QWP, HWP and TWP for all these three states. {In the experiment, MUB measurements are performed on $3\times10^6$ photons with two different settings in {Figs.~\ref{third_wave_plate_experimental_setup}(a) and \ref{third_wave_plate_experimental_setup}~(b)}. We first measure the state with the well-calibrated setting and assume the estimated state $\rho$ as the real state. Then we intentionally mis-calibrate the optic axes of the wave plates with an angular error, and obtain an estimation $\hat{\rho}$. Thus, the systematic error due to this angular error is calculated as $\text{tr}(\hat{\rho}-\rho)^2$. The estimated states by the well-calibrated TWP setting have a fidelity of over $99.9\%$ with those by the well-calibrated QWP-HWP setting for all the three states above, validating each other. In terms of systematic error,} experimental results (dots) and the theoretical results (solid lines) are shown in Fig.~\ref{error_1_qubit}, and they match very well. The performance of these two settings depend on $\rho$ and neither always outperforms the other. For example, for states in Figs.~\ref{error_1_qubit}(a) and \ref{error_1_qubit}(c), TWP beats the QWP-HWP combination while reversely for the state in Fig.~\ref{error_1_qubit}(b). However, there are more states where TWP performs better. The total systematic error in the estimation of these three states in the TWP setting adds up to be about two times smaller than that in the QWP-HWP setting as shown in Fig.~\ref{error_1_qubit}(d).

\subsection{Two-qubit tomography experiments}

\begin{figure}
\center{\includegraphics[scale=0.5]{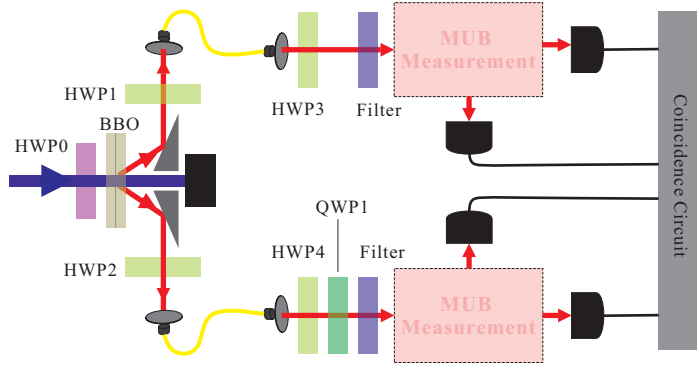}}
\caption{\label{two_third_wave_plate_experimental_setup}{Experimental setup for two-qubit tomography. A singlet state is prepared via SPDC process with a fidelity of 98\%. In quantum state tomography, single-qubit MUB measurements are implemented on both photons with TWP or the combination of QWP and HWP in Fig{s}.~\ref{third_wave_plate_experimental_setup}~(a) and \ref{third_wave_plate_experimental_setup}~(b). Coincidence events are recorded by a coincidence circuit.}}
\end{figure}

\begin{figure}
\center{\includegraphics[scale=0.55]{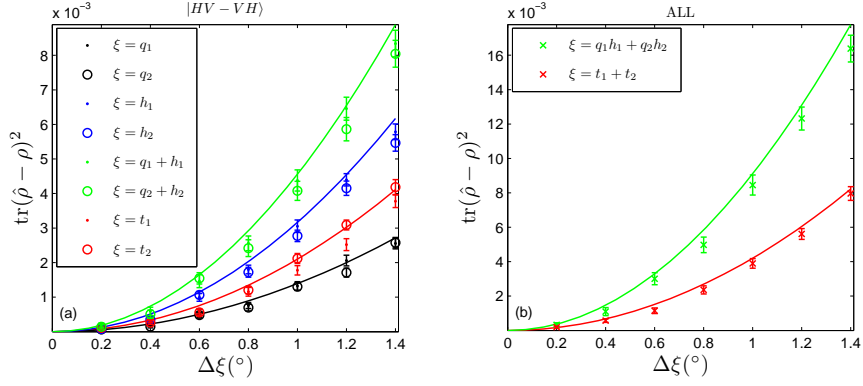}}
\caption{\label{error_2_qubit}{Systematic error in TWP and QWP-HWP setting for two-qubit tomography. In (a), systematic errors due to angle errors of QWP (black), HWP (blue), both of them (green) and TWP (red) are numerically simulated and denoted as solid lines. {The experimental results due to angle errors of wave plates are plotted as dots for photon 1 and circles for photon 2. The total systematic error for the two photons is plotted in (b)}. From (b), The TWP setting beats the QWP-HWP setting by about a factor of two. Error bars are the standard deviation of 100 Monte Carlo simulations with multinomial distribution of the counting statistics.}}
\end{figure}

In Fig.~\ref{two_third_wave_plate_experimental_setup}, a 100 mW, H-polarized beam at 404 nm from a continuous laser pumps a pair of type I phase-matched $\beta$-barium borate (BBO) crystals whose optic axes are normal to each other. After the spontaneous parametric down-conversion (SPDC) process, a pair of 808 nm photons are created. When the optic axis of half-wave plate (HWP0) at 404 nm is deviated $22.5^\circ$ from horizontal direction, the twin SPDC photons are maximally entangled. HWP1 and HWP2 rotate H and V to the fast and slow axes of the single mode fibers. At the output ports of the fibers, HWP3 and HWP4 rotate the polarization direction back to horizontal and vertical. QWP1 is tilted to compensate the phase of the entangled states to a singlet state. In the MUB measurement part, a complete set of MUB on either photon is performed with two methods: TWP setting in Fig.~\ref{third_wave_plate_experimental_setup}(a) and QWP-HWP setting in Fig.~\ref{third_wave_plate_experimental_setup}(b).

In the two-qubit case, the singlet state {$\rho=\ket{\Psi^-}\bra{\Psi^-}$} is chosen for three reasons: firstly, systematic error in the estimation of product states is a direct sum of that of single-qubit states; secondly, entangled states reveal the peculiar features of quantum systems and are valuable quantum resources; the last reason is that the systematic error in the estimation of Werner states is proportionate to the sum of the systematic error in the realization of single-qubit MUB for either photon \cite{hou_prepare}, which is similar to the systematic error averaged over unitarily equivalent states. The systematic error for Werner states \cite{hou_prepare} is
\begin{equation}\label{EQ:systematic error werner}
\text{tr}(\rho-\hat{\rho})^2=\frac{1}{4}p^2(\epsilon^2_1+\epsilon^2_2),
\end{equation}
where $\rho=p\ket{\Psi^-}\bra{\Psi^-}+(1-p){I}/{4}$, $\ket{\Psi^-}=\frac{1}{\sqrt{2}}(\ket{HV}-\ket{VH})$ and $\epsilon^2_i$ is the systematic error of the realized single-qubit MUB on photon $i$, $i=1,2$.

Similar to the qubit tomography experiment, we only experimentally measure the dependence of the systematic error on the angle errors of wave plates in the branches of both photons. Then product measurements of MUB are performed on $9\times10^5$ pairs of prepared singlet states. {The estimated state by the well-calibrated TWP setting has a fidelity of over $99.8\%$ with that by the well-calibrated QWP-HWP setting, agreeing well with each other.} From \eqsref{EQ: systematic error-TWP}, (\ref{EQ:systematic error werner}) and (\ref{EQ: systematic errors-QH Pauli}), angle errors of wave plates for either photon theoretically contribute $8(\Delta{t})^2$ in the TWP setting and $12(\Delta h)^2+5(\Delta q)^2$ in the QWP-HWP setting. However, numerical results in Fig.~\ref{error_2_qubit} (a) show that angle errors only cause $10.5(\Delta h)^2+4.6(\Delta q)^2$ in the QWP-HWP setting and $6.9(\Delta{t})^2$ in the TWP setting. This gap is due to the positive semi-definite conditions of density matrices, which arises when states are singular. Both the experimental results and numerical results in Fig.~\ref{error_2_qubit} (b) show that the systematic error in the TWP setting is only about a half of that in the QWP-HWP setting. The experimental results are slightly smaller than the numerical results because the prepared state is not exactly the expected singlet state, which only has a fidelity of $98\%$.

\section{Conclusion}\label{section:conclusion}

We have found that one wave plate is sufficient to realize two MUB as long as its phase is within $[45^\circ, 315^\circ]$. It is capable of realizing a complete set of MUB if the phase is within $[111.5^\circ, 141.7^\circ]$ or the symmetric interval about $180^\circ$. The systematic error in the realization of MUB in one wave plate setting is calculated to be twice smaller than that in the conventional QWP-HWP setting. TWPs are applied to single-qubit and two-qubit quantum state tomography experiments and experimentally show an error reduction by $50\%$ compared with the QWP-HWP combination. Other applications of TWP and arbitrary phase plates in the realization of any SU(2) and polarization state transformations need to be explored in the future.

\appendix
\section{Theoretical solutions at intersections}
\subsection{Theoretical solutions at the intersection around $\delta=120^\circ$}\label{section:APPEND-120}
From the numerical solutions in Fig.~\ref{all_angles}, some solutions cross at about $\delta=120^\circ$. Here solutions at this intersection are analytically calculated and the phase at this intersection is indeed $120^\circ$.
We denote the three red solid lines from bottom to top as $\theta_i, i=1,2,3$. The green, blue and black solid lines correspond to $90-\theta_i,90+\theta_i$ and $180-\theta_i, i=1,2,3$. Numerical solutions in Fig.~\ref{all_angles} show that crossing lines at the intersection should have the same coordinates, i.e.,
\begin{equation}\label{symmetry:120}
\begin{aligned}
90^\circ-\theta_1=90^\circ+\theta_1,\quad
\theta_2=\theta_3-90^\circ.
\end{aligned}
\end{equation}
From Eq.~(\ref{symmetry:120}), $\theta_1=0^\circ$. Substituting $\theta_1$ into Eq.~(\ref{bloch vector r}), one obtains $\vect{r}(\theta_1)=(0,0,1)^T$. With these relations, the conditions that $\vect{r}(\theta_i), i=1,2,3$ should be orthogonal to each other give two independent equations
\begin{equation}\label{mub condition:120}
\begin{aligned}
\cos \delta\sin^2 2\theta_2+\cos^2 2\theta_2=0,\quad
\sin^2\delta\sin^2 2\theta_2=\frac{1}{2}.
\end{aligned}
\end{equation}
The solutions of Eq.~(\ref{mub condition:120}) are $\delta=120^\circ, \theta_2=\frac{1}{4}\text{arccos}(-\frac{1}{3})\approx 27.37^\circ$ and $\theta_3=117.37^\circ$. Other solutions at this intersection represented by green, blue and black can be calculated from their symmetries about the red lines.

\subsection{Theoretical solutions at intersections around $\delta=126.3^\circ$ and $141.7^\circ$}\label{section:APPEND-126.3 and 141.7}
Numerical solutions in Fig.~\ref{all_angles} also show some solutions intersect near $\delta=126.3^\circ$ and $141.7^\circ$. Here we analytically calculate the solutions and phases at the intersections.
The three red lines are denoted as $\theta_i, i=1,2,3$ from bottom to top. In Fig.~\ref{all_angles} numerical solutions at the intersection around $\delta=141.7^\circ$ give
\begin{equation}\label{symmetry:end}
\begin{aligned}
\theta_1+\theta_2=90^\circ, \quad
\theta_3-90^\circ=180^\circ-\theta_3.
\end{aligned}
\end{equation}
From Eq.~(\ref{symmetry:end}), $\theta_3=135^\circ$ for the intersection around $\delta=141.7^\circ$. For the intersection at about $\delta=126.3^\circ$, one has
\begin{equation}\label{symmetry:end class two}
\begin{aligned}
\theta_1+\theta_3=90^\circ,\quad
90^\circ-\theta_2=\theta_2.
\end{aligned}
\end{equation}
Eq.~(\ref{symmetry:end class two}) gives $\theta_2=45^\circ$ for the intersection around $\delta=126.3^\circ$.
With Eqs.~(\ref{symmetry:end}) and (\ref{symmetry:end class two}), conditions of a complete set of MUB at the two intersections give the same equations
\begin{equation}\label{mub condition:end}
\begin{aligned}
\cos^2\delta\sin^2 2\theta_1+\cos \delta\cos^2 2\theta_1-\sin^2\delta\sin 2\theta_1=0,\\
(1-\cos \delta)^2\sin^2 2\theta_1\cos^2 2\theta_1=\sin^2\delta\sin^2 2\theta_1+(\cos \delta\sin^2 2\theta_1+\cos^2 2\theta_1)^2.
\end{aligned}
\end{equation}
The two equations in Eq.~(\ref{mub condition:end}) are equivalent to
\begin{equation}\label{mub condition:simplified end}
\begin{aligned}
x(x-1)y^2+x=(1-x^2)y,\\
(x-1)^2y^4=(x-1)^2y^2-\frac{1}{2}
\end{aligned}
\end{equation}
with $x=\cos \delta$ and $y=\sin 2\theta_1$. The square of the first equation in Eq.~(\ref{mub condition:simplified end}) is
\begin{equation}\label{mub condition:square}
\begin{aligned}
x^2(x-1)^2y^4+2x^2(x-1)y^2+x^2=(x-1)^2(x+1)^2y^2.
\end{aligned}
\end{equation}
Replacing $(x-1)^2y^4$ in Eq.~(\ref{mub condition:square}) with the second equation in Eq.~(\ref{mub condition:simplified end}), one obtains
\begin{equation}\label{mub condition:reduce parameter}
\begin{aligned}
y^2=\frac{x^2}{2(1-x^2)}.
\end{aligned}
\end{equation}
Substituting Eq.~(\ref{mub condition:reduce parameter}) into the second equation in Eq.~(\ref{mub condition:simplified end}), we have
\begin{equation}\label{x four order equation}
  3x^4+4x+2=0.
\end{equation}
This four-order equation has two real solutions $x=-0.5923$ and $x=-0.7854$, corresponding to $\delta=126.32^\circ$ and $141.76^\circ$, $\theta_1=15.66^\circ$ and $31.90^\circ$. From Eq.~(\ref{symmetry:end class two}), $\theta_3=74.34^\circ$ for $\delta=126.32^\circ$ and Eq.~(\ref{symmetry:end}) gives $\theta_2=58.10^\circ$ for $\delta=141.76^\circ$.

\section{Systematic error in the realization of MUB in the QWP-HWP setting}\label{section:APPEND-MUB systeamtic error QWP-HWP}
In the QWP-HWP setting, the realized bases are denoted as  $\vect{r}(\delta_q, \delta_h, q, h)$, where $\delta_q, \delta_h$ are the real phases of QWP and HWP in Fig.~\ref{configuration}(a). From \eqref{r:QWP-HWP} and \cite{error_compensation}, at $\delta_q=\frac{\pi}{2}$ and $\delta_h=\pi$,
\begin{equation}\label{EQ: r derivative QH}
\begin{aligned}
  \|\vect{r}_{\delta_q}\|^2&=\sin^2(4h-2q)=(r_2)^2,\quad  \|\vect{r}_{\delta_h}\|^2=\sin^2 2h,\\
  \|\vect{r}_{q}\|^2&={4+4}\cos ^{2}(4h-2q)={8-4}(r_2)^2,\quad  \|\vect{r}_{h}\|^2=16.
\end{aligned}
\end{equation}
From \eqsref{EQ:three MUB error} and (\ref{EQ: r derivative QH}), one obtains
\begin{equation*}
\begin{aligned}
({\epsilon_{\delta_q}})^2&=\sum_{j=1}^{3}(r_2^j)^2=1,\quad
({\epsilon_q})^2=\sum_{j=1}^{3}{8-4}(r_2^j)^2={20},\\
({\epsilon_h})^2&=48,\quad
({\epsilon_{\delta_h}})^2=\sum_{j=1}^{3}\sin^2 2h^j.
\end{aligned}
\end{equation*}
$({\epsilon_{\delta_h}})^2$ depends on the choices of MUB. In the realization of three Pauli operators with $h^1=h^2=22.5^\circ$ and $h^3=0^\circ$, $({\epsilon_{\delta_h}})^2=1$ and
\begin{equation}\label{EQ: systematic errors-QH Pauli}
\epsilon^2=48(\Delta h)^2+20(\Delta q)^2+(\Delta\delta_h)^2+(\Delta\delta_q)^2.
\end{equation}
For simplicity, $h^j$ is assumed to be uniformly distributed within $[0^\circ,360^\circ]$ in the consideration of general MUB. Then $({\epsilon_{\delta_h}})^2$ is averaged as $\frac{3}{2}$, and
\begin{equation}\label{EQ: systematic errors-QH}
\epsilon^2=48(\Delta h)^2+20(\Delta q)^2+1.5(\Delta\delta_h)^2+(\Delta\delta_q)^2.
\end{equation}

\section{Systematic error in qubit state estimation}\label{section:APPEND-qubit estimation systeamtic error}
From \cite{error_compensation,hou_prepare}, the systematic error in qubit state estimation with a complete set of MUB is
\begin{equation}\label{EQ: systematic error general}
\text{tr}(\rho-\hat{\rho})^2=\frac{1}{2}\sum_{\xi}\|\frac{\partial {R}^T}{\partial {\xi}}\vect{s}\|^2(\Delta {\xi})^2,
\end{equation}
where $\frac{\partial R}{\partial {\xi}}=(\frac{\partial \vect{r}^{(1)}}{\partial {\xi}},\frac{\partial \vect{r}^{(2)}}{\partial {\xi}},\frac{\partial \vect{r}^{(3)}}{\partial {\xi}})$ and $\vect{s}$ is the Bloch vector of $\rho$.

In the TWP setting with rotations angles at $0^\circ$, $27.37^\circ$ and $117.37^\circ$, from \eqsref{bloch vector r} and (\ref{r:TWP}),
\begin{equation}\label{EQ:R derivative-TWP}
\begin{aligned}
\frac{\partial R}{\partial{t}}&=\left(
                                                    \begin{array}{ccc}
                                                      3 & -1 & -1 \\
                                                      -\sqrt{3} & -1 & 1 \\
                                                      0 & -2\sqrt{2} & -2\sqrt{2} \\
                                                    \end{array}
                                                  \right),\quad
\frac{\partial R}{\partial{\delta_t}}&=\left(
                                       \begin{array}{ccc}
                                         0 & \frac{\sqrt{6}}{6} & \frac{\sqrt{6}}{6} \\
                                         0 & \frac{\sqrt{6}}{6} & -\frac{\sqrt{6}}{6} \\
                                         0 & -\frac{\sqrt{3}}{3} & -\frac{\sqrt{3}}{3} \\
                                       \end{array}
                                     \right).
\end{aligned}
\end{equation}
Thus, the systematic error $\text{tr}(\rho-\hat{\rho})^2$ in the estimation of the three states $\rho=p\ket{\phi}\bra{\phi}+(1-p){I}/{2}$ with $\ket{\phi}=\ket{H}, \ket{D}, \ket{R}$ is calculated from \eqsref{EQ: systematic error general} and (\ref{EQ:R derivative-TWP}) to be $[8(\Delta{t})^2+\frac{1}{3}(\Delta{\delta_t})^2]p^2$, $[\frac{11}{2}(\Delta{t})+\frac{1}{6}(\Delta{\delta_t})^2]p^2$ and $[\frac{5}{2}(\Delta{t})^2+\frac{1}{6}(\Delta{\delta_t})]p^2$.

In the QWP-HWP setting, from \cite{error_compensation}, the systematic error $\text{tr}(\rho-\hat{\rho})^2$ in the estimation of the three states can be calculated to be $[8(\Delta~h)^2+4(\Delta~q)+0.25(\Delta\delta_h)^2]p^2$, $[2(\Delta~q)+0.25(\Delta\delta_h)^2+(\Delta\delta_q)^2]p^2$ and $[16(\Delta~h)^2+4(\Delta~q)]p^2$.


\section*{Acknowledgments}
\addcontentsline{toc}{section}{Acknowledgments}%
The work was supported by National Fundamental Research Program (Grants No. 2011CBA00200 and No. 2011CB9211200), National Natural Science Foundation of China (Grants No. 61108009 and No. 61222504) and in part by the Australian Research Council's Discovery Projects funding scheme under Project DP130101658.
\end{document}